\documentclass[preprint]{elsarticle}
\usepackage[toc,page]{appendix}

\usepackage[colorlinks]{hyperref}
\hypersetup{
    colorlinks,
    citecolor=red,
    filecolor=cyan,
    linkcolor=blue,
    urlcolor=magenta
}
\usepackage{hyperref}

\usepackage{xcolor}
\usepackage{manfnt}
\usepackage{bm}
\usepackage{alltt,dsfont}
\usepackage{amsmath,amssymb}

\newcommand{\lessim} {\ {\raise-.5ex\hbox{$\buildrel<\over\sim$}}\ }

\newcommand{\gssim}{\ {\raise-.5ex\hbox{$\buildrel>\over\sim$}}\ }

\newcommand{\iden}{ \mathds{ 1}}

\newcommand{\tr}{{\text {Tr} }}
\newcommand{\si}{\sigma}
\newcommand{\sib}{\bar{\sigma}}
\newcommand{\tJ}{\ $t$-$J$ \ }
\newcommand{\beq}{\begin{eqnarray}}
\newcommand{\eeq}{\end{eqnarray}}
\newcommand{\barray}{\begin{eqnarray}}
\newcommand{\earray}{\end{eqnarray}}
\newcommand{\nn}{\nonumber}
\newcommand{\refdisp}[1]{Ref.~(\cite{#1})}

\newcommand{\half}{\frac{1}{2}}

\usepackage{xparse}

\newcommand{\disp}[1]{Eq.~(\ref{#1})}

\renewcommand{\k}{ {\vec{k}}}

\newcommand{\Ns}{N_{max}}

\newcommand{\bmat}{\begin{bmatrix}}
\newcommand{\emat}{\end{bmatrix}}
\newcommand{\bvmat}{\begin{vmatrix}}
\newcommand{\evmat}{\end{vmatrix}}

\makeatletter
\def\ps@pprintTitle{%
   \let\@oddhead\@empty
   \let\@evenhead\@empty
   \def\@oddfoot{\reset@font\hfil\thepage\hfil}
   \let\@evenfoot\@oddfoot
}
\makeatother 

\usepackage{atveryend}
\makeatletter
\let\origcitation\citation
\AtEndDocument{\def\mycites{}%
  \def\citation#1{\g@addto@macro\mycites{#1^^J}\origcitation{#1}}}
\AtVeryEndDocument{\newwrite\citeout\immediate\openout\citeout=\jobname.cit
 \immediate\write\citeout{\mycites}\immediate\closeout\citeout}
\makeatother

\begin{document}
\title{Fermi Surface Volume of   Interacting  Systems }
\author{B Sriram Shastry}
\ead{sriram@physics.ucsc.edu}
\address{Physics Department, University of California,  Santa Cruz, California 95064, USA }
\date{ \today}
\begin{abstract}
Three Fermion  sumrules for  interacting systems are derived at $T=0$, involving the number expectation $\bar{N}(\mu)$,  canonical chemical potentials $\mu(m)$,  a logarithmic time derivative of the Greens function $\gamma_{\k \si}$ and the static Greens function. In essence  we   establish at zero temperature the sumrules  linking: 
$$ \bar{N}(\mu) \leftrightarrow  \sum_{m} \Theta(\mu- \mu(m)) \leftrightarrow     \sum_{\vec{k},\si} \Theta\left(\gamma_{\vec{k} \si}\right) \leftrightarrow  \sum_{\vec{k},\si} \Theta\left(G_\si(\k,0)\right). $$
 Connecting them across leads to the        Luttinger and Ward sumrule, originally proved  perturbatively for  Fermi liquids. Our sumrules are nonperturbative in character and valid  in a considerably broader setting  that additionally  includes  non-canonical Fermions and Tomonaga-Luttinger models.    Generalizations are given for singlet-paired superconductors, where  one of the sumrules requires a testable  assumption  of particle-hole symmetry at all couplings. 
 The sumrules are found  by requiring  a {continuous} evolution from the Fermi gas, and  by assuming a    monotonic increase of  $\mu(m)$  with particle number $m$. 
 At finite T  a pseudo-Fermi surface,  accessible to angle resolved photoemission, is defined
using the zero crossings of the    first frequency moment of a weighted spectral function.  \end{abstract}

\maketitle

\section{Introduction \label{introduction} } The Luttinger-Ward (LW) sumrule  \cite{Luttinger-Ward} for interacting electrons   expresses the number of electrons in terms of  the static limit of the
 imaginary frequency Greens function\cite{Matsubara,AGD-paper,AGD}  for $T\to 0$  as
 \beq
\bar{N}(\mu)=  \sum_{\vec{k},\si} \Theta\left( G_\si(\k,\omega= 0|\mu) \right) \label{number-sumrule-LW},
\eeq
with $\Theta(x) = \half \left(1+ \mbox{sgn}(x) \right)$. Since the static Greens function  is negative outside  the Fermi surface, its volume is fixed by the number of particles \cite{Luttinger-Ward,Luttinger,AGD}, independent  of the magnitude  of the interaction. This interaction independence is a fundamental result  in Landau's theory of the  Fermi liquid\cite{Landau,Nozieres}. In  condensed matter physics,  field theory and statistical mechanics,  the  origin of this sumrule  and its ramifications  have been very influential\cite{deDominicis,Jona-Lasinio,Nozieres,AGD,Langer-Ambegaokar,Langreth}. It  has continued to receive much attention to recent times\cite{Altshuler,Dzyaloshinskii,Bedell,Phillips,Oshikawa,Yamanaka,Sachdev,others,Tsvelick,Hewson,Hewson2,Quinn,Seki-Yunoki}, partly motivated by the search for novel phases of matter that might violate this sumrule.
 The present work  provides a physically transparent derivation of the sumrule, and   extends it in several directions. The extended version includes  non-Fermi liquids, such as the  1-d Tomonaga Luttinger model (TLM).  It  is also valid  for non-canonical Fermions, such as U$=$$\infty$ Gutzwiller projected electrons in  the \tJ model, in treatments where  continuity with the Fermi gas is maintained \cite{ECFL}, but not otherwise  \cite{Quinn,ECQL}. Our extension also includes  singlet pairing superconductors. These include the s-wave BCS-Gor'kov-Nambu case and d-wave cuprate superconductors.  While   { analyticity} in the coupling is lost in these extensions,  they do evolve { continuously} from  the non-interacting  limit, which suffices for our purposes. Exotic superconductors, where one hollows out the k-space  \cite{Wilczek}, provide an interesting counter-point  where continuity with the gas limit is discarded;  we need to exclude them here too. 
 
Since the static Greens function entering \disp{number-sumrule-LW} is not directly measurable, 
 one  needs to relate it to other directly visible signatures for using it.  
 This work provides  a new and experimentally accessible sum-rule  \disp{neff}, which is equivalent to \disp{number-sumrule-LW} at T=0 in the cases considered. It also allows one to define a pseudo Fermi surface at any  T.  This surface carries  useful information on the real part of the on-shell self energy.

 \subsection{Methods used \label{methods}}
  The   technique used here  is non-perturbative, it relies  on  {\em isothermal} continuity in some parameter $\lambda$ connecting the interacting and non-interacting systems.   Since this type of continuity has not  been explicitly discussed in literature, a few words are in order.  As some parameter in the Hamiltonian is varied,  the  variation is required to be isothermal, i.e. at each intermediate value of the parameter, the system is allowed to  repopulate  energy levels according to the thermal distribution.  This is in contrast to adiabatic variations where the population of the energy levels is frozen at their starting values.
By continuity we imply that the expectation of the energy and other observables change without discontinuity, i.e. we rule out first order transitions. Illustrating the distinction  we note that the change of shape of the Fermi surface for anisotropic systems is allowed by isothermal continuity, but not by adiabatic continuity. Finally    our method does not require analyticity in a coupling, isothermal continuity is sufficient for our purpose. 

  Our other main   assumption is that the canonical ensemble (CE) chemical potentials $\mu(m)$ increase monotonically with the particle number $m$, whereby  the canonical free energy is thereby a concave-up function of $m$. This is tantamount to ruling out phase separation.
   We argue in Section.\ref{thermodynamics} that such a monotonic behavior could be regarded as a defining feature of   repulsive interactions.

   In each  case covered by our argument,  at non-zero $T$ we construct  an an effective particle density $n_{eff}(T)$, and pseudo-Fermi surface, whose temperature variation  reveals lowest lying characteristic energy scales in the system. The pseudo-Fermi surface  has the  potential to be  studied  using  angle resolved photo emission (ARPES) technique, and hence is discussed in some detail in Section (\ref{pseudoFS})

 \subsection{Organization of the paper \label{organize}}
The paper is organized   as follows.  
I  first establish in Section.\ref{thermodynamics} a  basic thermodynamic number sumrule  for  electrons with repulsive interactions;
\beq
\bar{N}(\mu) = \sum_{m=0}^{\Ns-1} \Theta(\mu- \mu(m)),  \label{number-sumrule} 
 \eeq
 where the CE chemical potential  $\mu(m)=F_{m+1}-F_m $ is the difference of the canonical free energies $F$ with $m+1$ and $m$ particles.   We will  assume a hard-core set of particles,   and therefore the maximum number of particles is limited by $\Ns$.  In Section.\ref{logderivative} I next introduce $\gamma$,  the {temporal}  log-derivative of the Greens function:
 \beq
 \gamma_{\k \si}(\mu,T)&=& \lim_{\tau=\beta/2} \partial_\tau \log G_\si(\k,\tau|\mu)  \label{gamma-def}, \ \;\;  \beta = \frac{1}{k_BT}.
\eeq
Setting $\tau=\beta/2$  sandwiches each Fermionic operator of $G$ symmetrically by factors that project all contributing states to the ground state as $T\to0$. While the study of $G_\si(\k,\frac{\beta}{2}|\mu)$ is  popular in quantum Monte-Carlo studies\cite{QMC}, the log-derivative, playing a key  role in this work, has not been discussed earlier. Its physical content at low $T$, as $\mu$ minus a $\vec{k}$-weighted  average  over $\mu(m)$ becomes clear later (see \disp{gamma-4}).  In the Section.\ref{logderivative} we  make an important  distinction between two ways of taking the   zero temperature and thermodynamic limits, in {\bf Limit-I}  we take T$\to$0 first and  $L \to \infty$ later,
while in {\bf Limit-II} we take  $L \to \infty$ first and   $T\to 0$ later.   

In Section.\ref{T=0}   the $T=0$ limit is taken first (i.e. in the limit {\bf Limit-I}), and shown to lead to the sumrule
 \beq
 \sum_{\vec{k},\si} \Theta\left(\gamma_{\vec{k} \si}(\mu,0)\right)= \sum_m \Theta(\mu- \mu(m)) \label{psi-sumrule-1}.
 \eeq
This is demonstrated for the Fermi liquid and also for the 1-d case of a Tomonaga-Luttinger model.

In Section.\ref{Normal1/L=0}  the $L\to \infty$ limit is taken first (i.e. in the limit {\bf Limit-II}), whereby we obtain a continuous frequency variable in terms of which a spectral function can be defined. Here the sumrule
\beq
 \sum_{\vec{k},\si} \Theta\left(\gamma_{\vec{k} \si}(\mu,0^+)\right)= \sum_{\vec{k},\si} \Theta\left(G_\si(\k,\omega=0|\mu)\right), \label{number-sumrule-zeroT}
\eeq
is established for Fermi liquids in Section.\ref{fermi-liquids} and for 1-d TLL systems in Section\ref{1dtll}.

 Assuming unbroken symmetry, powerful theorems on the  uniqueness of the ground state  \cite{yang-lee,lieb-lebowitz} are applicable, these allows us 
to   equate  the two zero temperature limits
 \beq \gamma_{\vec{k} \si}(\mu,0)=\gamma_{\vec{k} \si}(\mu,0^+).  \label{equality} \eeq 
   Upon using  Eqs.~(\ref{number-sumrule},\ref{Nsum-super})  Eqs.~(\ref{psi-sumrule-1},\ref{number-sumrule-zeroT}) (or Eqs.~(\ref{SC-sumrule},\ref{sczeroplus})) then  imply   the sumrule \disp{number-sumrule-LW}. In the infinite volume limit, the $\k$ sums are replaced by integrals as usual.  
    
   In Section.\ref{superconductor}  
 a systematic development of the volume theorem for a singlet superconducting state is provided.  This  broken symmetry state not accessible by  the methods  of L-W. 
 In Section.\ref{SCthermodynamics}  we study the canonical chemical potentials $\mu_e(2m) \equiv \half (F_{2m+2}-F_{2m})$ constrained to  the even particle sector. The $\mu_e(2m)$  are taken to be monotonically increasing in $m$, reflecting  the inherent repulsion between pairs of electrons. In this ensemble 
 we study the effects of adding or removing a particle and thence the Greens function, leading to the sumrule
 \beq
 \bar{N}_{SC}(\mu) =2 \sum_{m=0}^{\half\Ns} \Theta(\mu - \mu_e(2m)),  \label{Nsum-super}
 \eeq
which replaces \disp{number-sumrule} in the normal state.

In Section.\ref{SClimit1} the Greens function and $\gamma_\k$ are studied at T=0, (i.e. in the {\bf Limit-I})  in the superconducting state, subject to the assumptions of  particle hole symmetry and of the repulsion between the Cooper  pairs of electrons.  Here one finds 
\beq
\sum_{\vec{k}\si} \Theta(\gamma_{\vec{k} \si}(\mu,0)) = 2 \; \sum_m \Theta(\mu- \mu_e(2m)) = \bar{N}(\mu),  \label{SC-sumrule}
\eeq
 a sumrule corresponding to Eqs.~(\ref{psi-sumrule-1}).

In Section.\ref{SClimit2} the Greens function and $\gamma_\k$ are studied at L=$\infty$, (i.e. in the {\bf Limit-II})  in the superconducting state. Here we use the  Nambu-Go'rkov \cite{Nambu,Gorkov,Schrieffer,Tewordt,Schrieffer-Scalapino} formalism together with the formally exact quasiparticle representation\cite{Tewordt} of  the diagonal Greens function. This yields the sumrule \disp{sczeroplus}, and completes the set of links giving the number sumrule \disp{number-sumrule-LW}. In summary the sumrules corresponding to Eqs.~(\ref{psi-sumrule-1},\ref{number-sumrule-zeroT})  for a superconductor  are Eqs.~(\ref{SC-sumrule},\ref{sczeroplus}) in Section.\ref{SClimit1} and Section.\ref{SClimit2}.

In Section.\ref{pseudoFS} details of the applications of the sumrules at finite T to angle resolved photo emission (ARPES) are given. The main finding is that one can use  a first moment of the frequency with respect to the  weight function
\beq
{\cal W}(\k, \omega,T) ={W_0} \frac{A(\k,\omega)}{\cosh(\half \beta \omega)} \label{weight}
\eeq
where $W_0$ is a normalization constant and   $A(\k,\omega)$ is the electronic  spectral weight measured in experiments. It is denoted in the rest of the paper by the theoreticians favorite symbol $\rho_G(\k,\omega)$. The first moment  with respect to ${\cal W}$ of the frequency $\langle \omega \rangle_\k$ is found to be equal to  $- \gamma_\k(\mu,0^+)$  at T=0, and in view of the theorems proved here,  can be used as a proxy for the inverse static Greens function. It can be found from photoemission at any T,   and thereby  permits us to define an observable pseudo- Fermi surface (PFS), which becomes the true Fermi surface (FS) at $T\to 0$.  
 Section.\ref{PFSFL} examines  the T dependence of the pseudo FS and notes that it can be used to unravel the often sensitive  T dependence of the real part of self energy.  In Section.\ref{SC-calc} the pseudo FS for a singlet superconducting state is discussed in some detail.

In Section.\ref{discussion} I summarize the paper and discuss the results.

\subsection{The Hamiltonian} Consider  a  two component Fermion Hamiltonian 
\beq
{\cal H} = \sum_{\k \si} \varepsilon({\k})\, C^\dagger_{\k \si} C_{\k \si} + U\times \mbox{interaction} - \mu \, {\cal N} 
\eeq
\normalsize
in the grand canonical ensemble (GCE),
where ${\cal N}$ is the number operator,   $\mu$ is the (running i.e. varying) chemical potential, $\varepsilon(\k)$ the energy dispersion. {   We take  the interaction as  a short-ranged Hubbard type interaction,    possibly with a few further neighbor terms. The initial discussion assumes  $U>0$, and later we allow for pairing i.e. $U<0$.}
We  assume  a finite   lattice  in d-dimensions  with $N_s=L^d$ sites (L the linear dimension) and take the limit of an infinite system at the end. 

\section{ A   number  sumrule at T=0 \label{thermodynamics}}

We derive a new and  useful  sumrule \disp{number-sumrule} for the electron number at T=0 for electrons with repulsive interactions. It is of thermodynamic origin and is  based on an assumption of ``good behavior'' of the  chemical potentials of repulsive finite systems. Let us define the common eigenstates  of $  {\cal N}, {\cal H}$ as
$|m,a\rangle$
with   eigenvalues $ m, \, E_a(m) - m \mu$ as the respective eigenvalues. In the  canonical ensemble (CE)  $m$ particle sector, we will denote $E_0(m)$ and $F_m$ as the ground state energy and free energy $F_m= - k_B T \log Z_m$. 
 We define the CE  chemical potentials $\mu(m)$  using 
\beq
\mu(m) = F_{m+1}- F_m, \;\mbox{for}\;\; 0\leq m < \Ns, \label{chems}
\eeq
where $T$ dependence is implied in all variables.  The value of $\Ns$ is twice the number of sites  for the prototypical spin-$\half$  Hubbard model.  The  set of free energies $F_m$ is conveniently   extended  by defining $F_0$ and $F_{\Ns+1}$  satisfying the conditions   $F_0 > 2 F_1 - F_2$ and $F_{\Ns+1} > 2 F_{\Ns} - F_{\Ns-1}$ but are arbitrary otherwise. By inversion we obtain for $m \geq 1$
\beq F_m-F_0= \mu(m-1)+  \mu(m-2)+\ldots+\mu(0) \label{free-energies}.
 \eeq

 Our essential assumption is that of  a positive definite CE compressibility, i.e. a strictly concave-up  free energy, i.e. a strictly  increasing chemical potential for all $m$
\beq
F_{m+1}+ F_{m-1} - 2 F_m & >&  0, \nn \\
\mbox{or} \;\;\mu(m) >\mu(m-1) \label{concave-up}.
\eeq

 In a very large system, if we replace differences by derivatives, \disp{concave-up} becomes the more familiar condition of a positive physical  compressibility. We can use it  to order the CE  chemical potentials as a monotonically increasing set
 \beq 
 \mu(0) <\ldots  < \mu(j) < \ldots < \mu(\Ns). \label{ordered-mu}
 \eeq
From the interesting example of the Hubbard model on a buckyball cluster, we learn that  this  condition  can be   violated by  ostensibly repulsive   interactions\cite{kivelson}, leading to phase separation and related phenomena. Therefore the  ordering in  \disp{ordered-mu} seem to us to be  no more than  a robust   characterization of  truly repulsive interactions.

We introduce a useful set of weight functions
\beq
\xi_n= e^{\beta\left\{ \mu- \mu(n)\right\} }.
\eeq 
Using these  we may write $p_\mu(m)$  the probability of finding  $m$ particles in the GCE. With  $p_\mu(m)\equiv \exp{\beta (m \mu- F_m)}/Z(\mu)$, and the grand partition function  $Z(\mu) = \sum_m  e^{\beta (   \mu m- F_m)}$, we obtain 
\beq
    &&{p}_\mu(m)  =   Z^{-1}(\mu) \;\xi_0 \xi_1 \ldots \xi_{m-1} \\
 &&Z(\mu)=1 +\xi_0+\xi_0 \xi_1+\xi_0 \xi_1 \xi_2 +\xi_0 \xi_1 \xi_2\xi_3+ \ldots
  \label{pmu}
 \eeq
 The CE chemical potentials    $\mu(m)$ are computed at low $T$ from the ground state energies $E_0(m)$.

 When $T\ll 2\pi \hbar v/(L k_B)$ {where $v\sim v_F$ the band velocity,} the free energies $F_m$ can be replaced by the ground state energies $F_m \to E_0(m)$, and  
 the canonical chemical potentials    $\mu(m)$  computed  from the ground state energies $E_0(m)$.  We note that
  \beq
 \lim_{T\to 0} \xi_j \to   \begin{cases}
      \infty, & \text{if}\ \mu > \mu(j),  \\
      1, & \text{if}\  \mu= \mu(j), \\
      0, & \text{if}\ \mu < \mu(j).
    \end{cases}  \label{limiting}
    \eeq  Let us consider the case when $\mu$ is in the $j^{th}$ (open) interval ${\cal I}_j$ defined as
\beq
{\cal I}_j = \{ \mu \; | \;  \mu(j-1) < \mu  < \mu(j) \}. 
\eeq
When $\mu \in {\cal I}_j$ at very low $T$, the $j$ particle sector is  occupied  while  $j+1$ and higher sectors are unoccupied.  To see this, 
when $T\to0$  we observe that  $\xi_0,\xi_1\ldots \xi_{j-1}$ grow while $\xi_{j},\xi_{j+1}\ldots$ decrease towards zero. Therefore for $\mu \in {\cal I}_j$, $Z$ is dominated by a single term
\beq 
Z& =& \xi_0 \xi_1\ldots \xi_{j-1} \times {\cal Y}  \\
{\cal Y}&=&  \left(1 + \frac{1}{\xi_{j-1}} + \frac{1}{\xi_{j-1} \xi_{j-2}} + \ldots + \xi_j+\xi_j \xi_{j+1}+\ldots\right) \nn \\
&\to & 1,  \nn
\eeq
and therefore
\beq
p_\mu(j)\to 1
\eeq
while the probabilities with lower and higher  indices vanish:
\beq
 && p_\mu(j-r)\to \frac{1}{\xi_{j-1}\xi_{j-2} \ldots \xi_{j-r}}\sim 0 \nn \\
 && p_\mu(j+r)\to \xi_j\ldots\xi_{j+r-1}\sim 0.
\eeq
 Therefore at $T=0$ it follows that the system has $j$, and  no more than $j$ particles, i.e.
 \beq
\lim_{T = 0}  p_\mu(j) =  \Theta( \mu- \mu({j-1}) ) - \Theta(\mu-  \mu({j})).\label{pmu1}
\eeq
The number of particles can be found  using \disp{pmu1} and  $\bar{N}(\mu)= \sum_{m=1}^{N_s} m \,  p_\mu(m)$.  Shifting the sum in one of the terms  and simplifying, we deduce the   $T=0$ thermodynamic number sumrule \disp{number-sumrule}.

 Note the crucial role played by  concavity of the free energy, it implies   a $1\leftrightarrow1$  relationship between $m$ and $\mu(m)$. This   rules out  double bends {\small \dbend}, i.e a non monotonic relation which prevents inversion. The assumed monotonicity allows the  relationship  to be   inverted,  yielding  $m$ from $\mu(m)$  uniquely and hence giving the  sum-rule. In order to deal with degeneracies of $\mu(m)$, usually arising from discrete symmetries (spin, parity, rotation,..), we relax the strictly increasing condition \disp{concave-up}  to the weaker
 \beq
 \mu(m) \geq \mu(m-1),
 \eeq
 we obtain a second form of the sumrule:
 \beq
  \bar{N}(\mu) = \sum'_m g_m \Theta(\mu- \mu(m)), \label{degen-sumrule}
 \eeq
where $g_m$ is the degeneracy of the particular $\mu(m)$, and 
the primed sum is over unequal  $\mu(m)$'s.

\section{Log-derivative of the Greens function \label{logderivative}}
The  log-derivative  in \disp{gamma-def} can be written as a ratio  \beq \gamma_{\k \si}={\beta_{\k \si}}/{\alpha_{\k \si}}, \label{gamma-T0}\eeq where 
\beq
 \alpha_{\vec{k} \si}(\mu,T) &=&   -G_\si(\k,\frac{\beta}{2}|\mu)  \label{alpha-def} \\
 \beta_{\vec{k} \si}(\mu,T)& =&   - \lim_{\tau\to \half \beta} \partial_\tau G_\si(\k,\tau|\mu) \label{beta-def}.
 \eeq
In terms of the convenient variable 
 $${\bf f}(m,a,b)\equiv e^{\beta \left( \mu (m +\half) -\half  (E_a(m)+ E_b(m+1)) \right)}/Z(\mu),$$ we find
\beq
 \alpha_{\vec{k} \si}(\mu,T)  =   \sum_{m,a,b} {\bf f}(m,a,b)  |\langle m,a | C_{\k \si}| m+1,b\rangle|^2, \;\; \label{alpha-def-2} 
 \eeq
 and
  \beq
\beta_{\vec{k} \si}(\mu,T) &=&  \sum_{m,a,b} {\bf f}(m,a,b) \left(\mu+E_a(m)-E_b(m+1) \right) \nn \\
&& |\langle  m,a| C_{\k \si}|m+1, b\rangle|^2     \label{alpha-beta-def} 
\eeq
These  spectral representations  imply that at low T {\em both initial and final states} are limited to their   ground states in their respective number sectors.  

 We  take the low temperature limit and the thermodynamic limit in two distinct ways, by comparing  $k_BT$ with an  energy scale $\Delta_E$ representing the excited state energy level separation in gapless systems:
\beq
\Delta_E\sim \frac{2 \pi \hbar { v}}{L},  \label{DeltaE}
\eeq
 {where $v$$\sim$$v_F$ the band velocity}.  We  distinguish between  two ways of taking the  limit
\begin{itemize}
\item {\bf Limit (I)\;:}   $  \Delta_E > k_BT \gssim 0$, \;\;   or equivalently \;  $\{\frac{1}{L} \to 0, T\to 0\}$
\item {\bf Limit (II):}   $k_BT > \Delta_E \gssim 0$, \;\; or  equivalently \; $\{ T\to 0,\frac{1}{L}\to 0\}$.
\end{itemize}
The two limits can be taken with different sets of tools,  ${\bf Limit (I)}$ leads to \disp{psi-sumrule-1}, and can be taken employing ideas and tools relevant to  finite size   systems, while ${\bf Limit (II)}$ leading to \disp{number-sumrule-zeroT} allows the use   of  electronic spectral  functions that are continuous functions of $\omega$. { The results of \cite{yang-lee,lieb-lebowitz} imply \disp{equality}, i.e.  that the two  limits coincide asymptotically. }

\section{Zero temperature  Limit (I), i.e. $\{\frac{1}{L} \to 0, T\to 0\}$ \label{T=0}}
In this section we consider Fermi liquids of TLM systems and take the  T=0 limit  
in  Eqs.~(\ref{gamma-T0},\ref{alpha-def-2},\ref{alpha-beta-def}).
Upon  taking the stated limit,  we project the sum over the intermediate states  $a,b$ to the ground state, and write $ e^{\beta (\mu m - E_0(m))} \to p_\mu(m)\times Z(\mu)$, whereby
\beq
\alpha_{\vec{k}\si} (\mu,0)&=& \sum_{m} \Phi_m(\vec{k}\si),  \nn \\
   \beta_{\vec{k}\si}(\mu,0) &=& \sum_{m}  \Phi_m(\vec{k}\si) \times \left\{\mu - \mu(m) \right\}, \nn \\
  \gamma_{\vec{k}\si}(\mu,0) &=& \mu - \sum_m \widetilde{\Phi}_m(\vec{k}\si)  \mu(m), \label{gamma-4}
 \eeq
 the normalized  weight function
 $\widetilde{\Phi}_m= {{\Phi}_m}/{\sum_r {\Phi}_r}$   is  normalized to unity $ \sum_m \widetilde{\Phi}_m=1$) and  its un-normalized counterpart   
\beq
{\Phi}_m(\vec{k}\si)= {  p_\mu(m)  e^{\half(\beta-\mu(m))} \;\, Z_{\si N_s}(\vec{k},m) }, \label{hat-Phi}\\
Z_{\si N_s}(\vec{k},m)=   |\langle m,0| C_{\k \si}| m+1, 0\rangle|^2. \label{Zk}
\eeq
$Z_{\si N_s}$  is the ground state CE quasiparticle weight   of a state with $m$ particles in $N_s$. 
In  \disp{gamma-4} by writing $\mu(m) \langle m,0| C_{\k \si}| m+1, 0\rangle= \langle m,0| [C_{\k \si},H]| m+1, 0\rangle  $  and evaluating the  kinetic piece explicitly, we obtain
\beq
\gamma_{\vec{k}\si}(\mu,0) =  \mu -\varepsilon(\k) - {\cal M}(\vec{k}\si, \mu), \;\;\;\;  \label{gamma-m-relation}\\
{\cal M}(\vec{k}\si, \mu) =   {\sum_m \frac{\widetilde{\Phi}_m(\vec{k} \si)}{ Z^{\half}(\vec{k},m)} \langle m,0| [C_{\k \si},V]| m+1, 0\rangle }. \;\; \;\; \label{Mk}  \eeq
We  require $Z$ in \disp{Zk}  to be non-zero at {\em finite} $N_s$, although it could vanish
as $N_s\to \infty$, in such as way that the normalized $\widetilde{\Phi}$ and ${\cal M}$ involving the ratios of $Z$-like objects remain non-zero.
Let us also observe that  ${\cal M} $ vanishes   on  turning off  interactions. We comment  on its  relation to the conventional Dyson self energy below after \disp{gamma-E-relation}.

We next argue that  \disp{gamma-4} implies \disp{psi-sumrule-1} provided  the interacting system is continuously connected to the gas limit.  For the strictly monotonic   case $\mu(m)<\mu(m+1)$, there is a  $1\leftrightarrow1$ map between the $\k \si$ and the $\mu(m)$, extending the obvious map in the gas. Hence 
$\widetilde{\Phi}_m=\delta_{m,m_0}$ for some  $m_0$, whereby $\gamma_{\k\si}= \mu-\mu(m_0)$.  Summing over all $\k\si$ leads to \disp{psi-sumrule-1}.
 This property of a sum over all $\k \si$ recovering the sum over all $\mu(m)$  follows from the   completeness of Fermi operators with the $\k \si$ labels, i.e. there is no state in the Hilbert space that is inaccessible  by a combination of these operators.
 
We might  relax strict monotonicity of $\mu(m)$ and allow for the merging  of a set of $\mu(m)$ at different $m$ with say $\mu(m_0)$.  In this case  $\widetilde{\Phi}(m)$ are non-zero for  the set of $m$ with non vanishing matrix elements in $Z_{\si N_s}(\k,m)$.
 Summing over these $m$'s  we again get  $\gamma_{\k\si}= \mu-\mu(m_0)$. Further summing over all $\k \si$ gives us back \disp{psi-sumrule-1}, with a suitable degeneracy factor, provided  we  use completeness of the sum.  We  verify completeness  in the noninteracting case (including shell type degeneracies) by using the representation \disp{gamma-m-relation}, with ${\cal M}=0$. In an  interacting theory this  completeness   requires invoking isothermal continuity.

\section{Zero temperature   Limit (II), i.e.  $\{ T\to 0,\frac{1}{L}\to 0\}$ \label{Normal1/L=0}}
 We now consider the log-derivative $\gamma_{\vec{k}}(\mu,T)$ for  Fermi liquids as well as   1-d TLM.
We are interested in calculating the $T\to 0^+$ limit of  $\gamma_\k(\mu,T)$  near  its root. 

In order to calculate $\alpha$ (\disp{alpha-def}) and $\beta$ (\disp{beta-def}) (dropping the explicit spin label below) we use the spectral function representations  for the time dependent  $G$  detailed in Appendix-A. We start with \disp{g-tau-rep} where we put $\tau = \half \beta$ so that 
 \beq
  \alpha_\k(\mu,T) &=&   \int_{-\infty}^\infty \frac{d \omega}{2 \cosh(\beta \omega/2) } \, \rho_G(\k,\omega), \;\;\;\label{alpha-2} \\
\beta_{\vec{k}}(\mu,T) &=& - \int_{-\infty}^\infty \frac{\omega \; d \omega}{2 \cosh(\beta \omega/2) } \, \rho_G({\vec{k}},\omega),\;\;\; \label{beta-2}\\
\gamma_\k(\mu,T) &=& {\beta_\k}/{\alpha_\k} \label{gamma-2}.
 \eeq
\subsection{ Fermi liquids \label{fermi-liquids}}
The spectral function in a Fermi liquid can be expressed  for low $T, |\omega| \ll T_F$ as a Lorentzian\cite{AGD} 
\beq
\rho_G(\vec{k},\omega) \sim \frac{Z({\vec{k}})}{\pi} \frac{\Gamma_k}{\Gamma_k^2+ (\omega- E(\vec{k},T))^2}, \label{quasiparticle-spectralfunction}
\eeq
where the quasiparticle weight  
\beq Z^{-1}(\vec{k})=1-   \partial_\omega {\Sigma}(\vec{k},\omega)\big\vert_{\omega\to 0}\label{grand-Z}, \eeq 
and the width of the peak $\Gamma_k=- Z(\k) \Sigma''(\k,0,T)$, these are implicitly  functions of $k,T,\mu$ etc. Note that  $\Gamma_k\sim T^2$ is  the standard Fermi liquid result for this object. The quasiparticle energy is defined as usual from the root of the nonlinear equation
\beq
E(\vec{k},T) =\varepsilon(\k) +\Sigma'(\k, E,T) -\mu(T), \label{qp-root}
\eeq
and $\Sigma'$ ($\Sigma''$) denotes  the real (imaginary) part of $\Sigma$.
From \disp{useful-integrals} and using the convenient symbol
  \beq 
  W(\k,T) = \half+ \frac{\Gamma_k+ i E(\k,T )}{2 \pi T} \label{W}
  \eeq 
we deduce that
\beq
\alpha_{\k}(\mu,T) &=& \frac{Z(\k)}{\pi} \Re e \, \xi(W) \nn \\
\beta_{\k}(\mu,T) &=& - E(\k,T ) \alpha_{\k}(\mu,T) -  \frac{Z(\k) \Gamma_k}{\pi}  \Im m \,   \xi(W)   \nn \\
\gamma_{\k}(\mu,T) &=& -E(\k,T) - \Gamma_k  \frac{ \Im m \xi(W)}{\Re e \xi(W)}. \label{gamma-FL}
\eeq
In the limit $T\to 0^+$ the width $\Gamma_k$ vanishes and we obtain
\beq
\gamma_{\k}(\mu,0^+)= -E(\k,0). \label{gamma-E-relation}
\eeq
Comparing the relations Eqs.~(\ref{qp-root},\ref{gamma-E-relation}) with Eqs.~(\ref{gamma-m-relation},\ref{Mk}), we observe that the variable ${\cal M}$ is essentially the self energy $\Sigma'$ from  the perspective of {\bf {Limit-I}}. A change in the sign of $\gamma_{\k}$ therefore occurs at the zero of $E(\k,0)$. Close to this root, i.e. with small $E$   \disp{qp-root} and the Dyson expression for the Greens function 
give us
\beq
\Theta(\gamma_{\k}(\mu,0^+)) = \Theta( G^{-1}(\k,\omega=0|\mu)), 
\eeq
and by replacing the static Greens function $G^{-1}\to G$ we obtain the sumrule \disp{number-sumrule-zeroT}. We return to these expressions later  in Section (\ref{pseudoFS}),
where we carry out a detailed analysis of the volume of the Fermi surface in  connection with ARPES experiments.
 
\subsection{  Non-Fermi liquids in 1-d \label{1dtll}}
 In this section we apply our method to the  case of Tomonaga-Luttinger systems. This is an extensively   studied area where many methods for exact solution are available\cite{Gogolin,Kluemper,Giamarchi}.
  In these systems the quasiparticle weight $Z$ vanishes, in parallel to the discussion for {\bf Limit (I)}  after  \disp{Mk}. We show below that  as $T\to 0$ in {\bf Limit(II)}, $\alpha_\k$ vanishes as well, and so does $\beta_\k$ in such a way that $\gamma_\k$ remains finite and switches sign at the true Fermi wave-vector.

For  the canonical example of a  spinless model, the  spectral function  is severaly constrained by the Lorentz invariance of the theory and conformal invariance of the effective 2-d classical theory at finite T \cite{Gogolin,NFL,Kluemper}. It can be expressed as a scaling function   valid at low $T, \omega$ and $\hat{k}$ defined as $\hat{k}= k- \zeta k_*$  near the left ($\zeta$=-1) and right ($\zeta$=1) Fermi points $\mp k_*$ \cite{Giamarchi,Gogolin,NFL}
\beq
\rho_G({k},\omega)= \frac{1}{T^{{\alpha_0}}} \sum_{\zeta=\pm1} {\cal F}(\frac{\omega - \zeta V \hat{k}}{T}), \label{scaling-function}
\eeq 
where $V$     the renormalized Fermion velocity is related to the bare Fermi velocity $V_F$ by a  non singular scaling factor, and we set $k_B=1$ in this section.  Here and in the following
 we should retain only one of the two terms of  the $\zeta$ sum, where $\hat{k}$ is small.
 Although we did not specify the value of $k_*$ yet, it 
 will turn out  that $k_*=k_F$ below thanks to the sumrule. The exponent $\alpha_0<1$,  both $\alpha,V$  depend on the interaction strength, and the positive definite scaling function is peaked at the origin. It satisfies
 ${\cal F}(0)=1$ and  ${\cal F}(\xi)\to 1/|\xi|^{\alpha_0}$ for $|\xi|\gg 1$. As $T\to 0^+$
  we   obtain
 \beq
 \rho_G(k,\omega) \sim \sum_{\zeta=\pm1} \frac{{\cal A}}{|\omega - \zeta V \hat{k}|^{\alpha_0}},
 \eeq
with   ${\cal A}>0$.
From the above spectral function and \disp{reg} we can calculate the Greens function near zero frequency close to the Fermi points with  $T\to0^+$ as
\beq
G(k,0|\mu) = - {\cal B}   \frac{\zeta V \hat{k}}{|V \hat{k}|^{\alpha_0+1}}, \label{gtlm}
\eeq
where ${\cal B}>0 $, and $\zeta=\pm1$ for the right and left Fermi points. 

We next calculate  \disp{alpha-2} and \disp{beta-2} using \disp{scaling-function}. The $\cosh(\half \beta \omega)$ factor in \disp{alpha-2} cuts off frequencies with $|\omega| > T$, and if we restrict $|V \hat{k}|\lessim  T $ as well, then the dimensionless argument of the scaling function ${\cal F}$ in \disp{scaling-function}  is  at most of ${\cal O}(1)$, and we get no contribution to the integrals from a  regime where ${\cal F}(\xi)\to 1/|\xi|^{\alpha_0}$ .  We can therefore reasonably replace
 ${\cal F}$ by a Lorentzian
 \beq
{\cal F}  \sim \frac{{\cal C} T}{\pi}  \frac{{\cal C} T}{({\cal C} T)^2+ (\omega - \zeta V \hat{k} )^2} ,
\eeq
where ${\cal C}$ is a positive constant. This  enables the convenience of  an  explicit evaluation of the integrals. If needed it can be supplanted by  a more lengthy and  tedious argument that avoids this replacement, giving the same  answer.

 We therefore use the results \disp{useful-integrals} and Eqs.~(\ref{rexi},\ref{imxi}) to explicitly perform the integrals and write down at  low T the  results   when $V\hat{k}$ is small;
\beq
\alpha_\k(\mu,T)&=&  \frac{{\cal C} }{\pi}\, T^{1-\alpha_0}\; \Re e \, \xi\left(\half+ \frac{{\cal C} T + i V \hat{k}}{2 \pi T} \right) \label{eq49} \\
\gamma_\k(\mu,T)&=& -  \zeta V \hat{k} -  {\cal C} T \frac{ \Im m  \, \xi\left(\half+ \frac{{\cal C} T + i \zeta  V \hat{k}}{2 \pi T} \right)}{ \Re e \, \xi\left(\half+ \frac{{\cal C} T + i V \hat{k}}{2 \pi T} \right)} \label{eq50}
\eeq
and  $\beta_\k(\mu,T) = \gamma_\k(\mu,T) \alpha_\k(\mu,T)$. Here  $\zeta=\pm1$ for the right and left Fermi points.  Note that these equations closely resemble \disp{gamma-FL}. At finite $T$ both terms in \disp{eq50} vanish when $V\hat{k}$ vanishes.
 As $T\to 0^+$  the second term in \disp{eq50} drops out identically, 
and we get
\beq
\gamma_k(\mu,0^+)= - \zeta V \hat{k}. \label{gamma-Ek-NFL}
\eeq
Comparing with the static Greens function \disp{gtlm} we obtain
 \beq
\lim_{T\to 0^+} \Theta( \gamma_{\vec{k}}(\mu,T)) &=&  \Theta\left(G(k,0|\mu)\right) 
\eeq
 we get the sumrule \disp{number-sumrule-zeroT}. 
 
  We note that  the vanishing of the quasiparticle weight $Z$ in this model is reflected in the vanishing of $\alpha_\k$ in \disp{eq49} at $T\to 0$.  Away from the Fermi points $\beta_\k$ also vanishes but their ratio $\gamma_\k$ in \disp{eq50} is finite. It follows that  
  \beq
 2 \sum_k  \Theta\left(G(k,0|\mu)\right) 
=  2 \sum_k  \Theta\left(G_0(k,0|\mu)\right) 
  \eeq
since  each side equals the number of particles and therefore equates  the Fermi diameter of the interacting and non interacting theories.  Therefore the unspecified $k_*$ can now be identified with the bare Fermi momentum $k_F$.

\section{ Sumrules in the singlet superconducting state \label{superconductor} }
The volume theorem can be generalized  to  singlet superconductors. 
 Our work is inspired by an observation regarding  Gor'kov's (diagonal) Greens function describing the superconducting state \cite{Gorkov} (see Eq.~(14))
\beq
G(\k,\omega)&=& \frac{1}{2 }  \frac{u^2(\k)}{\omega-E(\k) +i 0^+}+ \frac{1}{2 }  \frac{v^2(\k)}{\omega+E(\k) +i 0^+}, \label{Gorkov-g}\\
u^2(\k)&=&\half \left(1+ \frac{{\varepsilon(\k)-\mu}}{E(\k)}\right)\; \mbox{and} \;  v^2(\k)=\half \left(1- \frac{{\varepsilon(\k)-\mu}}{E(\k)}\right) \label{ukvk}.
\eeq
Here $\varepsilon(\k)$ is the band energy,  $E(\k)=\sqrt{\varepsilon^2(\k)+ \Delta^2(\k)}$ the (positive) quasiparticle energy and  $\Delta(\k)$ the gap function.
 It is remarkable to note that this expression contains in its innards,   a precise encoding  of the (submerged) normal  state Fermi surface. Setting $\omega=0$ we find
\beq
G(\k,0)=  \frac{\mu- \varepsilon(\k)}{E^2(\k)}. \label{Gorkov-static}
\eeq
Therefore in   system exhibiting superconductivity,  the submerged normal  state Fermi surface is revealed by a change in sign of  $G(\k,0)$ occurring at 
\beq
\varepsilon(\k)=\mu \label{simple-Gorkov},
\eeq
 and  at the root,
\beq
u(\k_F)=v(\k_F) \label{phsymmetry}.
\eeq
 The latter condition
 expresses an emergent  particle hole symmetry on the Fermi surface of the weak coupling  BCS solution. 
   While the above relations are  true at the mean-field (BCS) level of description, it is not clear if this encoding  survives the effects of strong   interactions, and further    refinements of the theory. It is also not entirely clear as to how one might probe this encoding, since there is  no known method for probing the static $G$ directly.   The first question is treated here with an affirmative answer for a fully gapless superconductor. For a partially   or fully  gapped case, it is subject to the survival of the particle hole symmetry as at least an approximate symmetry  for arbitrary coupling. The second is answered in Section.\ref{pseudoFS}, where we relate an observable first moment of frequency with a suitable weight function to this observation, thereby suggesting a potentially useful photoemission experiment.

 The strategy used for the normal state is extended to the superconductors as follows.  We first establish the thermodynamic sumrule \disp{Nsum-super}, under  the assumption  that (Cooper) pairs of electrons exhibit mutual repulsion,  when viewed  as composite particles.  We then  take the T=0 limit to obtain the sumrule \disp{SC-sumrule}.  The main assumption here,  used without a direct proof, is that  of nearly   particle hole symmetric matrix elements for the interacting system, analogous to \disp{phsymmetry} for the free case. Finally we take the $L\to \infty$ limit and using results from the Nambu-Gorkov formalism, obtain  \disp{sczeroplus} and hence the sumrule \disp{number-sumrule-zeroT}. This completes the set of sumrules needed to establish \disp{number-sumrule-LW} for the superconducting state as well.

\subsection{Superconducting phase: Thermodynamic sumrule \label{SCthermodynamics}}
We next study the thermodynamic sumrule for a superconducting state, using the canonical ensemble. This approach is familiar from the nuclear physics context \cite{Migdal,Bohr-Mottleson-Pines} and has been recently applied in the context of mesoscopic superconductivity \cite{Ambegaokar}. 
Our treatment initially assumes a finite gap such as  s-wave BCS superconductors, and later  generalized to include d-wave case relevant for cuprate superconductors. We define the CE chemical potentials remaining  within the {\em even-canonical}  ensemble\cite{Migdal,Ambegaokar}:
\beq
\mu_e(2m) \equiv \half (F_{2m+2}-F_{2m}), \label{chem-def}
\eeq
 and  require  the monotonic  property $ \mu_e(2m+2) > \mu_e(2m)$ leading to an ordering
 \beq
\mu_e(0) <\mu_e(2)< \ldots \mu_e(2j) \ldots < \mu_e(N_{max})  \label{pair-chem}. 
\eeq
This clearly implies  a concavity condition  on the free energies $2 F_{2n} < F_{2n-2} + F_{2n+2}$, arising from represents repulsion between pairs, 
 so that further clusters of electrons   are forbidden, i.e. {\em the pairing stops with pairs}. This results in a homogeneous many-body eigenstate  of pairs,   qualitatively  similar and continuously connected to a gas of (repulsive) Bosons as envisaged in \refdisp{Leggett,Nozieres-Schmitt-Rink}.  Assuming the ordering  \disp{pair-chem} we may repeat the discussion leading   to \disp{number-sumrule}, yielding \disp{Nsum-super}, the number sum-rule for paired superconductors at $T=0$, with the  extra factor of 2  from  skipping  odd fillings. 

\subsection{Superconducting phase: T=0 sumrule \label{SClimit1}}

We now consider the low T log-derivative Greens function  as in \disp{gamma-def} with  $F_m\sim E_0(m)$.
For this we also need the odd sector  energies $F_{2n+1}$, these  are expressed  in terms of  the even  energies and a gap function $\Delta$\cite{Migdal,Bohr-Mottleson-Pines,Richardson,Tinkham,Ambegaokar,Gaudin,vonDelft,Mateev,Amico,Amico2,Caux,Emil}:
 \beq
 E_0({2n+1})=\half( E_0({2n})+E_0({2n+2})) + \Delta(2n+1). \;\; \;\;\;\; \label{gap}
 \eeq
Here $\Delta$ playing the role of the BCS gap is assumed non-zero initially. It is interpretable as the  energy of an unpaired electron in an otherwise paired state. It  is essentially the lowest energy Bogoliubov-Valatin\cite{B-V}  quasiparticle in the CE.

 We consider the {\bf Limit(I)} of $\alpha,\beta$ of Eqs.~(11,12,13,14) of the main paper. We proceed to calculate Eq.~(13,14) by grouping pairs of terms  $\{2m,2m+1\}$, rewriting $p_\mu$'s  using Eq.~(27) and Eq.~(29) as $p_\mu(2m) = e^{\beta( 2m \mu - F_{2m})}/Z(\mu)$ and
 $p_\mu(2m+1) = e^{- \beta \Delta(2m +1)} {p^{\half}_\mu(2m) p^{\half}_\mu(2m+2)}$.  We further define the  matrix elements: 
 \beq V^{ab}_{\vec{k} \si}(2m+1)&=& \langle 2m+1,a| C_{ \vec{k}\si}|2 m+2,b \rangle, \nn \\
 \; U^{ab}_{\vec{k} \si}(2m+1)&=& \langle 2m,a | C_{\vec{k}\si}|2 m+1,b \rangle. \label{UV}
 \eeq
 The ground state $|2m+1,0\rangle$ has one unpaired quasiparticle (with 2-fold degeneracy), while $|2m, 0\rangle$ and $|2m+2,0\rangle$ are the  fully paired non-degenerate ground states. These matrix elements are therefore analogs of the familiar GCE coefficients $(u_k,v_k)=\sqrt{\half(1\pm\frac{\xi_k}{E_k})}$ of the BCS-Gor'kov theory\cite{BCS,B-V,Gorkov}
 noted in \disp{ukvk}, with $\xi = \varepsilon - \mu$. Recall that $\xi_k=0$ at the Fermi momentum, therefore the relation $u_{k_F}=v_{k_F}$  noted in \disp{phsymmetry} holds good\,  in weak coupling\cite{BCS,B-V,Gorkov}.  This relation  also underlies  the   Majorana zero modes   discussed in \cite{Majorana}, and is often viewed as expressing an  emergent particle-hole symmetry. Following this  we assume  the more general  ground states matrix elements satisfy $|U^{00}_{\k \si}| \sim |V^{00}_{\k \si}| $,   for the correct bridging momentum.   For finite systems  we require it to hold within a tolerance that is discussed below.
 
We closely follow the procedure in the Fermi liquid case, and express $\alpha,\beta$ in terms of the matrix elements $U,V$. 
\beq
\beta_{\vec{k}}(\mu,T) =\sum_{m} e^{ \half \beta( \mu - \mu_e(2m) - \Delta)} {p^{\half}_\mu(2m) p^{\half}_\mu(2m+2)} \; {\cal B}(m), \;\;\nn  \label{SC-beta}
\eeq
and
\beq
\alpha_{\vec{k}}(\mu,T)= \sum_{m} e^{ \half \beta( \mu - \mu_e(2m) - \Delta)} {p^{\half}_\mu(2m) p^{\half}_\mu(2m+2)} \; {\cal A}(m), 
\eeq
where
\beq
{\cal A}(m)= \left\{(U^{00}_{k \si}(2m+1))^2+ e^{-\beta (\mu - \mu_e(2m))} (V^{00}_{k \si}(2m+1))^2\right\},\label{SC-alpha-2}
\eeq
\beq
&&{\cal B}(m)=(\mu - \mu_e(2m) ) {\cal A}(m)+ \nn \\
 &&\;\; \Delta(2m+1) \left\{ (U^{00}_{k \si}(2m+1))^2- e^{-\beta (\mu - \mu_e(2m))} (V^{00}_{k \si}(2m+1))^2 \right\}.\;\; \label{SC-beta-2}
\eeq
In computing $\gamma_{\vec{k}}(\mu,T)$ as $T \to 0$, our calculation proceeds similar to the non-superconducting case but with the role of $Z(\vec{k},m)$ now played by the matrix elements $U,V$. Assuming continuity from the Fermi gas via the weak coupling BCS-Gor'kov theory, the given wave vector $\vec{k}$ picks out a single particle number $m$ contributing to both $\alpha,\beta$. In  the gapless case  for the given $\k$,   $\Delta$ vanishes as an inverse power of $L$. Thus 
${\cal B/A}= \mu- \mu_e(2m) $ with negligible corrections. If the   gap is non-zero
 the ratio ${\cal B/A}= \mu- \mu_e(2m) + \Delta \, {\cal C}$ where on dropping indices: $${\cal C}= \frac{ e^{\beta (\mu- \mu_e(2m))}  U^2- V^2}{e^{\beta (\mu- \mu_e(2m))} U^2+ V^2}.$$   
  We require the correction $ \Delta \, {\cal C}$ to be small  relative to the separation between $\mu_{e}(2m)$ and   $\mu_{e}(2m+2)$. 
If particle hole symmetry were exactly true then $U=V$,   
 $ \Delta \, {\cal C}=0$ 
and the node  in $\gamma$ is situated exactly at $\mu= \mu_{e}(2m)$ even if $\Delta \neq 0$. In practice  an approximate  equality between $U$ and $V$ suffices for this condition  with a specified tolerance. If we require that
\beq
\frac{|U^2-V^2|}{U^2+V^2} < \frac{|\mu_e(2m\pm 2)-\mu_e(2m)|}{\Delta(2m+1)},  \label{ph-exact}
\eeq
the node in $\gamma$ at $\mu \sim \mu_e(2m)$  is essentially unshifted.   Assuming this relation and summing over $\k$,  it follows  that
\beq
\sum_{\vec{k}\si} \Theta(\gamma_{\vec{k} \si}(\mu,0)) = 2 \; \sum_m \Theta(\mu- \mu_e(2m)) = \bar{N}(\mu), \nn \eeq
as noted in \disp{SC-sumrule},
where the factor of 2 comes from the equal contribution from $\vec{k}\si$ and its time reversed partner $-\vec{k} \sib$.

 

\subsection{Superconducting phase: $T\to0^+$   sumrule using  Nambu-Gor'kov formalism  \label{SClimit2}}

We next take the  limit $\{ T\to 0,\frac{1}{L}\to 0\}$ \label{1/L=0}  in the superconducting state. In order to go beyond the mean-field treatment in the  Gor'kov's paper\cite{Gorkov}, we use the formally exact Nambu formalism \cite{Nambu}. It contains all possible many body effects, including those neglected in mean field theory.  We start with  the  Nambu-Gor'kov \cite{Nambu,Gorkov,Schrieffer} four component  theory where the self energy  in the superconducting state
is expanded as
\beq
\Sigma(\k,z)= (1- Z_\k(z))z \; {\bf{\iden}}+ \phi(\k,z)\; {\bf {\tau}}_1 + \chi(\k,z)\; {\bf \tau}_3 \label{Nambu-selfenergy},
\eeq
 with $z=i \omega_n$ where the Nambu self energies $Z,\chi,\phi$ are even functions of $z$.
In this  notation for superconductors  $Z\sim 1-\partial_\omega \Sigma$, i.e  the inverse of the normal state convention where $Z\sim(1 -\partial_\omega \Sigma)^{-1}$.
 From this the
matrix  Greens function ${\bf G}$ is written as 
\beq
{\bf G}(\vec{k}, z|\mu)= \frac{z Z_\k(z) \iden + \tau_3( \varepsilon(\k) - \mu + \chi_\k(z)) + \tau_1 \phi_\k(z)}{z^2 Z^2_\k(z) - E^2_\k(z)}.\;\; \;\;\;\;\label{Nambu-Gorkov}
\eeq
We are only interested in the diagonal Greens function $G_{11}$, which we shall denote by $G$ below. This is the component of the Greens function relevant to the volume  theorem
and also to photoemission studies. It can be found  within the quasiparticle approximation  by expanding \disp{Nambu-Gorkov} near the poles of the Greens function   \cite{Nambu,Tewordt,Schrieffer, Schrieffer-Scalapino}. The poles of $G(\k,\omega)$ are located at the Bogoliubov-Valatin  (B-V)\cite{B-V} quasiparticle energies  $\omega = \pm E_{r \k}$ where
\beq
 E_{r \k}= \Re e (E_\k(\eta_\k)/Z_\k(\eta_\k)), \;\;\;\mbox{with}\;\;\; \eta_\k=E_{r \k}+ i 0^+,
\eeq 
and have a width $$\Gamma_\k= Z_\k^{-1} \Im  m \{ \eta_\k Z_\k(\eta_\k) - \frac{1}{E_\k} (\tilde{\varepsilon}_\k \chi_\k(\eta_\k) + \phi_\k \phi_\k(\eta_\k))    \},$$
expressed in terms of the following set of real constants  (Eq.~(2.25,2.26,2.27) of \cite{Tewordt}).
\beq
&&\tilde{\varepsilon}_\k= \varepsilon(\k)-\mu+ \Re e \, \chi_\k(\eta_\k), \; \phi_\k= \Re e \, \phi_\k(\eta_\k)\nn \\
&&E_\k= (\tilde{\varepsilon}^2_\k+ \phi_\k^2)^{\half}, \; Z_\k = \Re e Z_\k(\eta_\k).
\eeq
In the above expression $\phi_\k$ plays the role of a gap function, $\tilde{\varepsilon}_\k$ the  dispersion of a gapless underlying Fermi liquid renormalized  with self energy $\chi_\k$, and $E_\k$ is proportional to the quasiparticle energy $E_{r \k}$. 

   For energies close to the  BV quasiparticle energies, the quasiparticle  Greens function is given by the asymptotic expressions
\beq
G(\k, i\omega_n|\mu) &\sim& \sum_{\alpha=\pm1} \left\{\half+\alpha \frac{\tilde{\varepsilon}_\k}{2  E_{\k }} \right\} \frac{Z_\k^{-1}}{i\omega_n- \alpha E_{r \k } + i \Gamma_\k}, \;\;\;\;\;\;\; \label{GNambu}  \\
\rho_G(\k,\omega)&\sim& \frac{1}{\pi }\sum_{\alpha=\pm1} \left\{\half+\alpha \frac{\tilde{\varepsilon}_\k}{2  E_{\k }} \right\}  \frac{Z_\k^{-1} \Gamma_\k}{(\omega- \alpha E_{r \k } )^2+ \Gamma^2_\k} \label{spectral-Nambu}
\eeq
The spectral function  has a similar  status for superconducting systems as \disp{quasiparticle-spectralfunction}
 for Fermi liquids; both expressions capture the various many-body renormalizations in terms of a few  parameters. 
 
 We calculate $\alpha_\k,\beta_k$ from Eqs.~(\ref{alpha-2},\ref{beta-2}) using the spectral function \disp{spectral-Nambu} in greater detail below in  \disp{SCGamma} in Section (\ref{SC-calc}). However as $T\to 0^+$ it is known \cite{Tewordt} that $\Gamma_\k\to 0$, i.e. one has sharp poles, and $\rho_G$ is a sum over two delta functions. In this case we easily calculate   
\beq
\alpha_\k\sim \frac{1}{Z_k \cosh(\half \beta E_{r \k})}, \; \beta_\k \sim \frac{- \tilde{\varepsilon}_\k}{Z^2_\k \cosh(\half \beta E_{r \k})}\;\; \nn
\eeq
 therefore $\gamma_{\k}(\mu,T)\to-\frac{\tilde{\varepsilon}_\k}{Z_\k}$. 
Now  $G(\k,0|\mu)=- \frac{\tilde{\varepsilon}_\k}{E^2_\k}$ from \disp{GNambu}, and therefore
\beq
\Theta\left(\gamma_{\k } (\mu,0^+)\right)= \Theta\left( G(\k,0|\mu)\right), \label{sczeroplus}
\eeq
and therefore by summing over $\k$ we obtain the sumrule \disp{number-sumrule-zeroT}.
This  result is  argued to be valid for all flavors  of singlet pairing, including the gapless d-wave case.  We combine \disp{sczeroplus} or \disp{number-sumrule-zeroT}  with \disp{SC-sumrule} and infer the  sumrule \disp{number-sumrule-LW} in the superconductor.

\section{ The pseudo Fermi surface at finite $T$ \label{pseudoFS}} 
 Extending the  ground state sum-rule  to
 finite $T$, we  define a ``pseudo-Fermi surface'' and an  effective density  $n_{eff}(T)$ 
from the  changes in sign with $\vec{k}$ of $\gamma_{\vec{k} \si}(\mu,T)$. These tend to  the true Fermi surface and  particle density when $T\to0$, and  can be extracted  from experimental photoemission  data as follows.
In terms of  a dipole matrix-element $M$ and the Fermi function $f(\omega)=(\exp{\beta \omega}+1)^{-1}$,  the photoelectron intensity is given by
${\cal I}(\k,\omega)=M(\k) \rho_G(\k,\omega) f(\omega)$. From Eqs.~(\ref{gamma-def},\ref{alpha-2},\ref{beta-2}) it follows that $\gamma$ is a suitably  weighted {\em first moment of frequency}:
 \beq
 \gamma_{\k \si}(\mu,T)  &=& - \langle \omega \rangle_\k, \label{firstmom}
 \eeq
 where
 \beq
\langle \omega \rangle_\k &=&   \int d\omega \; {\cal I}(\vec{k},\omega)   e^{\half \beta \omega} \omega  \Big/ \int d\omega \;  {\cal I}(\vec{k},\omega)  e^{\half \beta \omega}, \label{sumrule-0} \\
&=&  \int  \;\rho_G(\vec{k},\omega) \frac{ \omega \; d\omega}{\cosh(\half \beta \omega)}  \Big/ \int  \; \rho_G(\vec{k},\omega) \frac{  d\omega}{\cosh(\half \beta \omega)} ,
 \label{sumrule-1} \eeq 
 the two expressions  Eqs.~(\ref{sumrule-0}, \ref{sumrule-1}) are equivalent since the $\k$ dependent matrix element and other factors cancel out. This weight function was already mentioned in \disp{weight} in the Section.\ref{introduction}.
 In  averaging over $\omega$, the  weight factors  provide exponential  cutoffs for high $|\omega|$. By replacing $\omega$ by $\omega^m$ in \disp{firstmom}, one can  generates the $m^{th}$ moment  $\langle \omega^{m} \rangle_\k$. This novel set  of moments characterize the  low energy excitations of the spectral function, unlike the moments without the T dependent weight functions,  and seem promising for further study.
 
  From $\gamma$ we define the effective density
   \beq n_{eff}(T)={1}/{N_s} \sum_{\k\si}\Theta
\left( \gamma_{\k \si}(\mu,T)\right). \label{neff} \eeq
We can  now define  the pseudo-Fermi surface at any T; it is defined as the set of Fermi points $\k$ satisfying
\beq
\langle \omega \rangle_\k=0. \label{pseudo-FS}
\eeq
The  sign changes of $\gamma$ with $\vec{k}$ occur on this surface, and $n_{eff}(T)$ counts the number of particles within this surface from \disp{neff}. It reduces to the true Fermi-surface at $T$=0.    We next discuss the content of this sum-rule at finite T for two important cases.

\subsection{Finite T volume sumrule:  Fermi liquids \label{PFSFL}}
We note that Eqs.~(\ref{firstmom},\ref{sumrule-0}) are identical to \disp{gamma-2} in Section(\ref{1/L=0}).
Therefore for Fermi liquids  at finite (but low)  T,  we can use the 
quasiparticle approximation for the spectral function \disp{quasiparticle-spectralfunction},
 so that
\beq
\langle \omega \rangle_\k= E(\k,T) + \Gamma_k  \frac{ \Im m \xi(\half+ \frac{\Gamma_k+ i E(\k,T )}{2 \pi T})}{\Re e \xi(\half+ \frac{\Gamma_k+ i E(\k,T )}{2 \pi T})},\label{firstmom-2}
\eeq
following \disp{gamma-FL}.  In order to deduce the pseudo Fermi points,  we observe that when $E(\k,T)$ vanishes in \disp{gamma-2}, the imaginary part of  of $\xi$ vanishes as well.    Thus at any T the pseudo Fermi point is located by
\beq
E(\k,T) = 0, \label{eq-a}
\eeq
 where $E(\k,T)$ is defined in \disp{qp-root}. At $T=0$  it reduces to
\beq
E(\k_F,0)= \varepsilon(\k_F) + \Sigma'(\k_F, 0,0)- \mu(0) =0, \label{eq-b}
\eeq
where we set  $\k=\k_F$, the corresponding T=0 Fermi momentum upon using   the volume theorem.  In  \disp{qp-root}   we  expand the self energy at low $\omega$ and write  $E$ as
\beq
&&Z^{-1}(\k,T) E(\k,T)= \varepsilon(\k) + \Sigma'(\k,0,T) - \mu(T) \label{QP-energy} \\
&=& \varepsilon(\k) + (\Sigma'(\k,0,T)-\Sigma'(\k,0,0)) + \Sigma'(\k,0,0)+(\mu(0) - \mu(T))-\mu(0) \nn \\
\eeq
The vanishing of the  right hand side locates the pseudo FS. Using \disp{eq-b}  we obtain 
\beq
(\mu(T) - \mu(0))&=&(\varepsilon(\k) -\varepsilon(\k_F))+
 (\Sigma'(\k,0,T)-\Sigma'(\k,0,0)) \nn \\
 && +(\Sigma'(\k,0,0)-\Sigma'(\k_F,0,0)).  \label{eq-c}
\eeq
As expected this equation is satisfied identically  by setting $\k=\k_F$ at $T=0$. At low T we perturb   by expanding $\k$ about $\k_F$,  
\beq
\k=\k_F+\delta \k,
\eeq
and   linearize in $\delta \k$  to find
\beq
\delta \k . \vec{V}_{\k_F} = (\mu(T) - \mu(0))- (\Sigma'(\k_F,0,T)-\Sigma'(\k_F,0,0)), \label{deltak-1}
\eeq
where $\vec{V}_\k= \vec{\nabla}_\k [\varepsilon(\k)+ \Sigma'(\k,0,0)]$ is the Fermi velocity. 
The variation $\delta \k$ is normal to the  true (i.e. T=0) FS, and can be determined from this relation.  Proceeding further  we may  write the change in  FS area  with T as a line integral over the  true  FS
\beq
\delta A(T)= \oint_{FS} dk_\perp  \frac{(\mu(T) - \mu(0))- (\Sigma'(\k_F,0,T)-\Sigma'(\k_F,0,0))}{|\vec{V}_k|}, \label{delta-A}
\eeq
where $dk_\perp$ is the wave-vector element  tangential to the FS. 

 The effective density at $T$  differs from the true particle density by the usual  counting rules leading to  
\beq n_{eff}(T) - n = 2\times \delta A(T)/(2 \pi)^2. \label{delta-n}
\eeq
The variation \disp{delta-A} is driven by the T dependent shifts of $\mu(T)$ and of  the real part of the self energy $\Sigma'(k_F,0,T)$. The shift of $\mu$ with T is the smaller of the two, and  can in principle be estimated experimentally. For example in ARPES  the apparent change of excitation energy with T of some fixed (T independent) feature, such as a band edge can be used for this purpose. The variation $\delta A(T)$ is amplified  when the quasiparticle  Fermi velocity is reduced from the bare one, as it often happens in strongly correlated matter.  An example of the  $T$ dependence of $n_{eff}$  in the \tJ model is shown in Ref.~\cite{Mai-Shastry}, where the variation with T is quite significant due to strong correlations.  revealing  emergent  low-energy scales in the problem.

The expression \disp{delta-A} allows us to explore the T dependent shift of the real part of $\Sigma$. This object is of great interest in strongly correlated materials.  In the strange metal regime of the d=$\infty$ Hubbard model, it has been reported in  \refdisp{second-order} (Fig (12.c))  to have a strong T dependence, which in turn leads to a  linear T resistivity

\subsection{Finite T  volume sumrule:  Superconductors  \label{SC-calc}}
In parallel to the treatment of the normal case above, we calculate the first moment \disp{sumrule-1} in the superconducting phase, using the quasiparticle spectral function in \disp{spectral-Nambu}, and the useful integrals noted in \disp{useful-integrals}. Cancelling common factors we write
\beq
&&\langle \omega\rangle_\k = {\cal N}/{\cal D}, \nn \\
{\cal N}&=& \sum_{\alpha=\pm1} \left\{\half+\alpha \frac{\tilde{\varepsilon}_\k}{2  E_{\k }} \right\}
\left[\alpha E_{ r \k}\,  \Re e \, \xi \left(\half+ \frac{\Gamma_\k+ i \alpha E_{r \k}}{2 \pi T}\right)+ \Gamma_\k \, \Im m \, \xi \left(\half+ \frac{\Gamma_\k+ i \alpha E_{r \k}}{2 \pi T}\right) \right]\nn \\
{\cal D}&=& \sum_{\alpha=\pm1} \left\{\half+\alpha \frac{\tilde{\varepsilon}_\k}{2  E_{\k }} \right\}  \Re e \, \xi \left(\half+ \frac{\Gamma_\k+ i \alpha E_{r \k}}{2 \pi T}\right) \label{temp-1}
\eeq
We now use the  properties of the $\xi$ function \disp{rexi} and \disp{imxi} to simplify  \disp{temp-1}. This gives the final formula for the first moment:
\beq
\langle \omega\rangle_\k= \frac{\tilde{\varepsilon}_\k}{  E_{\k }} \left( E_{r \k} + \Gamma_\k \frac{\Im m \,  \xi \left(\half+ \frac{\Gamma_\k+ i  E_{r \k}}{2 \pi T}\right)}{\Re e \, \xi \left(\half+ \frac{\Gamma_\k+ i  E_{r \k}}{2 \pi T}\right)} \right). \label{SCGamma} 
\eeq
The $\Gamma_\k$ term in the  above expression  is expected to be exponentially small for the gapped states and a power law for gapless singlet paired states.
This expression resembles \disp{firstmom-2} for the Fermi liquid state with  the energy dispersion $\tilde{\varepsilon}_\k $ replacing the quasiparticle energy $E(\k,T) $. The vanishing of the first moment locates the pseudo-FS for the superconductor  through the condition
\beq
\tilde{\varepsilon}_\k= \varepsilon(\k) - \mu + \Re e \, \chi_\k(E_{r \k}) \to 0, \label{Gorkov-renorm}
\eeq
which replaces the simple relation of Gor'kov's  theory \disp{simple-Gorkov}.  This implies that the shift in the chemical potential  from the noninteracting value due to  pairing effects  is exactly cancelled by the self energy term $ \Re e \, \chi_\k(E_{r \k})$.  This cancellation is  analogous to exact cancellation in the normal state.

 Our treatment of the pseudo FS of the superconducting state has a few  precedents.  The  closely related papers \cite{Gros,Randeria}  discuss the
surface formed by $\k$ with  $G(\k,0)=0$,  using a phenomenological model of $G$ for superconductors in strongly correlated cuprate materials. The model uses a  ``renormalized'' mean field theory\cite{RMFT} for this calculation. This method incorporates  effects of strong correlations through a rescaled version  of the BCS effective Hamiltonian with density dependent scale factors. The area of the above surface in these works   is found to be only approximately the number density, even at T=0. Their results are in contrast to the findings of the present work, where the pseudo FS area must match the particle density exactly at T=0. The discrepancy could be due  to missing a cancellation between the shifts of the self energy and  the  chemical potential, or due to  a lack of the (unproven) particle-hole symmetry at strong coupling.
Experimental checks of the particle-hole symmetry, as suggested in this work would be of considerable interest.

In the present work we 
propose a new  suggestion for probing the pseudo- Fermi surface for superconductors.
It differs from the signatures proposed earlier \cite{Gros,Randeria},  advocating either locating the maxima of the spectral weight, or the maxima of the gradient of the  momentum distribution function  $|\nabla_k n_k|$. Our proposal involves  studying the 
 first moment of the frequency $\langle \omega\rangle_\k$, defined in \disp{sumrule-1}. Its vanishing as in \disp{pseudo-FS} defines the pseudo FS.
As explained  above this  moment can be constructed from the dynamical information in ARPES.    For  singlet superconductors such as the cuprates,  the   pseudo-Fermi surface is definable on both sides of the superconducting transition using the  moment in both phases Eqs.~(\ref{firstmom-2},\ref{SCGamma}).  Apart from (usually  small) T dependent corrections, its area is the same in both phases, being related to  the  number of particles.

\section{ Discussion \label{discussion}}

Given the   importance of the Fermi volume theorem, and the attendant  complexities of deriving it,  a fresh  approach seems relevant.
This work presents a  non-perturbative derivation of the volume sum-rule \disp{number-sumrule-LW} in a broad setting.  
 We avoid  using the traditional number sum-rule 
 \beq
 \bar{N}(\mu)= \sum_{k \si} G_\si(\k,\tau=0^-|\mu);
 \eeq
  instead  we use different ways to compute the zero $T$  limit of the log-derivative $\gamma_\k$. This is  a major departure from the L-W route,  where the introduction of the Luttinger-Ward functional is  an essential second step. This functional 
can only be defined in perturbation theory, and  leads to difficulties for strong coupling problems, as explicitly demonstrated  in recent work \cite{Kozik}.

 In 1-d  \refdisp{Yamanaka} uses  adiabatic evolution of the  system with a magnetic flux parameter, to give a non-perturbative argument  for the invariance of Fermi diameter.  \refdisp{Oshikawa}
extends this to arbitrary dimensions d$>$1  assuming that  the system is a Fermi liquid.  In contrast to  the 1-d result,  \refdisp{Oshikawa}  also requires adiabatic evolution through a large  number   ${\cal{O}}(L^{d-1})$ of level crossings, arising from a large accumulation of phase with increasing flux.

More generally the use of adiabatic theorem in gapless situations, particularly  for d$>$1 is risky, and often requires extra symmetry for justification.  A well known example is provided by  a  gapless metal for   d$>$1, with a varying interaction strength.    When the symmetry is less than circular (or spherical),    k-space redistribution always occurs upon varying the interaction constant. This   results in  a change of shape of the Fermi surface \cite{Nozieres,Pomeranchuk,Metzner},   implying  that  adiabaticity is violated.   

  We note a few points regarding perturbative arguments.
The T=0 Brueckner-Gammel-Goldstone formalism\cite{BG} is based on the adiabatic theorem and  uses the  non-interacting Greens function $G_0$ as the foundation for the perturbation expansion. Therefore the invariance of the Fermi volume, as well as its shape, are automatic byproducts, we get back what we    initially put in. 
A  critique of this method by Kohn and Luttinger\cite{Kohn-Luttinger} led to the   L-W work. They  used finite T perturbation theory instead,  allowing for a k-space redistribution of occupied states\cite{Nozieres,Pomeranchuk,Metzner}.
 However   the   problem of strong coupling   remains.  It is hard to see how the L-W method can be justified in strong coupling,   recalling that  it is  predicated on the existence  of  the    Luttinger-Ward functional,    defined term by term  in  powers of the coupling. Recent work explicitly displays    pathologies of the L-W functional in Hubbard type models at strong coupling\cite{Kozik}.

  The present work utilizes   continuity, instead of   perturbation, to bypass the strong coupling problem.  Isothermal continuity breaks down at  level crossing transitions with tuning, and is signaled   by a jump in expectation values. Therefore
the guarantor of isothermal  continuity is the absence of jump discontinuities in expectation values. In summary   we may   assume isothermal continuity  within a continuously connected phase of matter, thus requiring  the absence of first order quantum  transitions. As our example of the superconductor shows, the isothermal argument  works through the normal to superconducting  transition, where the dependence on coupling  is non-analytic (but continuous). Here the adiabatic methods seem to fail.

 We use continuity in a parameter for linking the interacting system   with the Fermi gas.   The parameter used is most often, but not necessarily,   the coupling constant in the Hamiltonian. In the case of the \tJ model with extreme coupling  $U=\infty$, a more general interaction type parameter $\lambda  \in[0,1]$  is  invoked\cite{ECFL}.  Continuous evolution with $\lambda$ ensures the volume theorem for the \tJ model\cite{ECFL}.

Our extension of the volume theorem to singlet superconductor is based on two assumptions. Firstly we assume that pairs of electrons act repulsively with respect to other pairs, thereby giving a monotonically increasing chemical potential $\mu_{e}(m)$ in \disp{pair-chem}. This is certainly true in the BCS theory and in exactly solvable models\cite{Richardson,Gaudin} for  superconductivity in  finite size systems. It would break down if an as yet undiscovered    glue were to result in say  Cooper-quartets, instead  of Cooper-pairs.
The other main assumption is that of an approximately valid particle hole symmetry \disp{ph-exact} for the case of fully or partially gapped superconductors. This  leads to U$\sim$V in the correlated superconductors, extending the  known result   \disp{phsymmetry} in the weak coupling  BCS-Gor'kov case. This symmetry  has been assumed to be   true in  other  contexts, e.g. for the recently discussed Majorana Fermions \cite{Majorana}. For strongly coupled systems, 
this  symmetry  is  hard to establish analytically. However numerical tests of the condition \disp{ph-exact} involving ground-state to ground state matrix elements of the Fermi operators may be   feasible for small systems using exact diagonalization, and are planned for future work.  Finally since we establish a direct connection with observable variables,  one  could test the resulting  sumrule experimentally in a  variety of superconducting materials.  The results  would indicate if  this symmetry  holds good, and  how widely, if so.

The present work leads to the notion of a pseudo-Fermi surface defined finite T. This surface is shown here to be accessible to ARPES studies from moments of the observed intensities. It
  seems well worth  exploring this object and its T dependence experimentally to throw light on 
 interesting issues in strongly correlated matter. For superconductors such  measurements could complement information from the   high magnetic field setups used to study the same submerged normal  state Fermi surface by destroying the superconducting order using strong magnetic fields\cite{Greg}.

\section{Acknowledgments}
I am grateful to P. W. Anderson, P. Coleman, B. Doucot,   A.  Georges, A. C. Hewson,  H. R. Krishnamurthy, E. Perepelitsky,  M. Randeria,  and A. Tsvelik for helpful discussions  on   aspects of this problem  at various times. 
  The work at UCSC was supported by the US Department of Energy (DOE), Office of Science, Basic Energy Sciences
(BES), under Award No. DE-FG02-06ER46319.

\appendix

\section{ Spectral function and its relation to the Greens function \label{Appendix-A}}
With $-\beta < \tau \leq \beta$, 
we recall the (Matsubara) imaginary time Greens function \cite{Matsubara,AGD,AGD-paper} 
\beq G_\si(k,\tau) = - \frac{1}{Z(\mu)} \tr \, e^{- \beta {\cal H}} \left(T_\tau C_{k \si}(\tau) C^\dagger_{k \si}\right), \label{G}
\eeq 
where the time dependence is    $Q(\tau) = e^{\tau {\cal H}} Q e^{-\tau {\cal H}}$. Using the usual antiperiodicity $G(\tau)=-G(\tau+\beta)$ we define the Fourier version as usual $G(i\omega_n)= \half \int_{-\beta}^\beta G(\tau) e^{i \omega_n \tau} d\tau$.
We may express $G$  as
 \beq
 G(\vec{k},i \omega_n) &=& \int d\omega \, \frac{\rho_G(\vec{k},\omega)}{i\omega_n - \omega},\;\
\eeq
where the spectral function $\rho_G(\vec{k},\omega)$  can be conveniently  found from the analytic continuations $i \omega_n \to z$ followed by $z \to \omega+ i 0^+$ as
$ \rho_G(\vec{k},\omega) = - \frac{1}{\pi} \, \Im m G(\vec{k}, \omega + i 0^+)$.

  The spectral function  has a further representation\cite{AGD} 
\beq
\rho_{G}(k,\omega) &=&(1+e^{-\beta \omega})
 \sum_{n,m,a,b} p_\mu(n)  |\langle n, a | C_{ \vec{k}} | m, b \rangle |^2  \nn \\ 
&& \times \delta(\omega+ {{ E}}_{a}(n)-  {{ E}}_{b}(m)+ \mu) 
e^{- \beta   E_{a}(n)+\beta F_n}, \eeq
where $F_n$ is the n-particle free energy.

In terms of $\Delta_E\sim |E_a(n)-E_{a'}(n)|$, i.e.  a typical excitation energy at a fixed number for a finite system,
we may distinguish between two regimes. At zero T, or more generally  for $\Delta_E/k_B\gssim T$, the spectral function is a sum over separated delta functions and hence is very grainy. On the other hand provided $(T,\omega)\gssim \Delta_E/k_B$,  the sum over the delta functions is taken over several states and hence the resulting spectral functions are smooth functions of $\omega$. This is therefore a complementary regime to the earlier one.

 In terms of the spectral functions we may write the time dependent functions as
\beq
G(\vec{k},\tau) = \int_{-\infty}^\infty {d \omega}\, \rho_G(\vec{k},\omega)  e^{- \tau \omega} \left( f(\omega) \theta(-\tau)  -\bar{f}(\omega) \theta(\tau) \right) \nn \\
\label{g-tau-rep}
\eeq
 with the Fermi functions  $f(\omega)=\frac{1}{e^{\beta \omega}+1}$ and $\bar{f}=1-f$.  We will need the following relation for the real part of the Greens function
\beq
G(k,0) = -{\cal P} \int \frac{ d\omega}{ \omega} \rho_G(k,\omega), \label{reg}
\eeq
where ${\cal P}$ denotes the principal value.

 \section{Some useful integrals arising in the sum-rule \disp{sumrule-1} \label{Appendix-B}}
We outline the calculation of integrals that arise in \disp{sumrule-1}:
\beq
{\cal J}_m = \frac{1}{2 \pi} \int_{-\infty}^\infty  \; d \omega\; \frac{(-\omega)^m}{\cosh(\half \beta \omega)} \frac{\Gamma}{\Gamma^2 +(\omega- E)^2}, 
\eeq
for real parameters $\Gamma,E$
with $m=0,1$.  A simple way to do these integrals is to use the Mittag-Leffler expansion
\beq
\frac{1}{\cosh(\pi z)} = \frac{2}{\pi} \sum_{n=0}^\infty \frac{(-1)^n (n +\half)}{z^2+ (n +\half)^2},
\eeq
so that we can integrate term by term using the simple result for convolution of two Lorentzians. This yields
\beq
{\cal J}_0&=& \frac{1}{\pi} \sum_{n=0}^\infty (-1)^n \frac{\Gamma/(2 \pi T)+ n + \half}{E^2/(2 \pi T)^2+(\Gamma/(2 \pi T)+ n + \half)^2} \nn \\
{\cal J}_1&=& -  \frac{1}{\pi} \sum_{n=0}^\infty (-1)^n \frac{ (n + \half)E}{E^2/(2 \pi T)^2+(\Gamma/(2 \pi T)+ n + \half)^2}. \label{req-sums}
\eeq
These  sums can be performed using the digamma function
\beq
\Psi(z) = \frac{d}{d z} \log \Gamma(z) = \lim_{M\to \infty} \left(\log M - \sum_{n=0}^M \frac{1}{z+n} \right).
\eeq
We define  a meromorphic function $\xi(z)$  via
 the alternating infinite  sum
\beq
\xi(z)= \sum_{n=0}^\infty \frac{(-1)^n}{z+n} = \half \left(\Psi(\half+\frac{z}{2}) - \Psi(\frac{z}{2}). \right),\label{xi}
\eeq
In the complex $z$ plane $\xi(z)$
  has a pole at the origin and at every negative integers, and  is analytic  everywhere else. Writing $z=x+ i y$ we record  the useful corollaries
\beq
\Re e \, \xi(x+ i y) &=& \sum_{n=0}^\infty (-1)^n \frac{x+n}{(x+n)^2+y^2} \nn \\
\Im m \, \xi(x+ i y) &=& - \sum_{n=0}^\infty (-1)^n \frac{y}{(x+n)^2+y^2}.
\eeq

Using these we can perform  the required summations in  \disp{req-sums} analytically as
\beq
{\cal J}_0&=& \frac{1}{\pi} \Re e \, \xi \left( \half+ \frac{\Gamma+ i E}{2 \pi T} \right) \nn \\
{\cal J}_1&=& - \frac{E}{\pi} \Re e \, \xi \left( \half+ \frac{\Gamma+ i E}{2 \pi T}\right)  - \frac{\Gamma}{\pi} \Im  m \, \xi \left( \half+ \frac{\Gamma+ i E}{2 \pi T} \right). \label{useful-integrals}
\eeq
From the series defining  $\xi(z)$ in  \disp{xi}, it  is real for real $z$. 
Using the Schwarz reflection principle  we deduce relations needed in the text;  
for $\alpha = \pm 1$
\beq
\Re e \, \xi \left(\half+ \frac{\Gamma+ i \alpha E}{2 \pi T}\right)&=&\Re e\,  \xi \left(\half+ \frac{\Gamma+ i  E}{2 \pi T}\right) \label{rexi} \\
\Im m  \, \xi \left(\half+ \frac{\Gamma+ i \alpha E}{2 \pi T}\right)&=&\alpha \, \Im m  \, \xi \left(\half+ \frac{\Gamma+ i  E}{2 \pi T}\right). \label{imxi}
\eeq

\end{document}